\magnification1200
\input pictex.tex
%\vsize=195truemm
%\hsize=130truemm
%\parindent0.5truecm
%\parskip0pt
\parindent0pt
\parskip3pt minus3pt

\baselineskip=1.5\baselineskip

\def\plik{dwa$\_$dwumianowe$\_$estymacja$\_$roznica$\_$ver2a}
\def\dzien{\sl \number\day/\number\month/\number\year}
\footline{\hss{\sl \plik\ \dzien;\ \folio}}
\footline{\hss{\sl \folio}}
%\nopagenumbers

\def\tr{{\hat\vartheta}}
\def\m{\phantom{-}}

\centerline{\bf A New Exact Confidence Interval}
\centerline{\bf for the Difference of Two Binomial Proportions}

\vskip1truecm
\centerline{\bf Wojciech Zieliñski\footnote{}{{\sevenrm Results of the paper were partially obtained in cooperation with Department of the Prevention of Environmental Hazards and Allergology, Medical University of Warsaw, Banacha 1a, 02-097 Warszawa, Poland}}}
\bigskip
\centerline{Department of Econometrics and Statistics}
\centerline{Warsaw University of Life Sciences}
\centerline{Nowoursynowska 159, 02-776 Warszawa, Poland}
\centerline{e-mail: wojciech\_zielinski@sggw.pl}
\centerline{http://wojtek.zielinski.statystyka.info}
\vskip1truecm

\bigskip
{\bf Abstract} We consider interval estimation of the difference between two binomial proportions. Several methods of constructing such an interval are known. Unfortunately those confidence intervals have poor coverage probability: it is significantly smaller than the nominal confidence level. In this paper a new confidence interval is proposed. The construction needs only information on sample sizes and sample difference between proportions. The coverage probability of the proposed confidence interval is at least the nominal confidence level. The new confidence interval is illustrated by a medical example.
\bigskip

\bigskip


\vfill\eject

\bigskip
{\bf 1. Introduction}

Let $\xi_1$ and $\xi_2$ be two independent r.v.'s distributed as $Bin(n_1,\theta_1)$ and $Bin(n_2,\theta_2)$, respectively. We estimate the difference between the probabilities of success, i.e. $\vartheta=\theta_1-\theta_2$. Construction of confidence intervals for the difference of proportions has a very long history and has been widely studied, due to its numerous applications in biostatistics and elsewhere; see e.g. Anbar (1983), Newcomb (1988), Zhou et al. (2004). In all those constructions, normal approximation to the binomial distribution is applied. As a consequence it may be observed that the coverage probabilities of the asymptotic confidence intervals are less than the nominal confidence level (for a single binomial proportion see for example Brown et.al 2001). This is in contradiction to Neyman's (1934, p. 562) definition of a confidence interval. In what follows, a new confidence interval is proposed. That confidence interval is based on the exact distribution of the difference of the observed numbers of successes. A similar method was applied in constructing a confidence interval for a linear combination of proportions (W. Zieliñski 2018).

The paper is organized as follows. In the second section  a new confidence interval is constructed. In the third section a medical example is discussed. Some remarks and conclusions are collected in the last section.
In the first appendix there is given a short R-project program for calculating proposed confidence intervals. In the second appendix some known confidence intervals for the difference of probabilities are cited.

\bigskip
\goodbreak
{\bf 2. A new confidence interval}
\bigskip

Let $\xi_1\sim Bin(n_1,\theta_1)$ and $\xi_2\sim Bin(n_2,\theta_2)$ be independent binomially distributed random variables. The random variable $\tr={\xi_1\over n_1}-{\xi_2\over n_2}$ is the minimum variance unbiased estimator of $\vartheta=\theta_1-\theta_2$.

The confidence intervals widely used in applications are constructed in the following statistical model:
$$\left(\{0,1,\ldots,n_1\}\times\{0,1,\ldots,n_2\},\left\{Bin(n_1,\theta_1)\cdot Bin(n_2,\theta_2), 0\leq\theta_1,\theta_2\leq1\right\}\right).$$
Since we are interested in estimating $\vartheta=\theta_1-\theta_2$ on the basis of $\tr$, we consider the new statistical model
$$\left({\cal X},\left\{{\cal P}(n_1,n_2,\vartheta), -1\leq\vartheta\leq1\right\}\right),$$
where
$${\cal X}=\left\{{k_1\over n_1}-{k_2\over n_2}: k_1\in\{0,1,\ldots,n_1\},k_2\in\{0,1,\ldots,n_2\}\right\}.$$
The family $\left\{{\cal P}(n_1,n_2,\vartheta), -1\leq\vartheta\leq1\right\}$ of distributions  is as follows. Since for a given $\vartheta\in(-1,1)$ the probability $\theta_1$ is a number from the interval $(a(\vartheta),b(\vartheta))$, where
$$a(\vartheta)=\max\{0,\vartheta\}\quad\hbox{and}\quad b(\vartheta)=\min\{1,1+\vartheta\},$$
the probability of the event $\{\tr=u\}$ (for $u\in{\cal X}$) equals (simply apply the law of total probability and averaging with respect to $\theta_1$)
$$\eqalign{
P_\vartheta\{\tr= u\}&=P_\vartheta\left\{{\xi_1\over n_1}-{\xi_2\over n_2}= u\right\}\cr
&={1\over L(\vartheta)}\int_{a(\vartheta)}^{b(\vartheta)}\sum_{i_2=0}^{n_2}Q_{(\theta_1,n_1)}\left\{\xi_1= {n_1}\left(u+{i_2\over n_2}\right)\right\}Q_{(\theta_1-\vartheta,n_2)}\left\{\xi_2=i_2\right\}d\theta_1.\cr}$$
Here $L(\vartheta)=b(\vartheta)-a(\vartheta)$ and $Q_{(\mu,m)}\left\{\zeta=k\right\}={m\choose k}\mu^k(1-\mu)^{m-k}$ for $k=0,1,\dots,m$.

Note that the family $\left\{{\cal P}(n_1,n_2,\vartheta), -1\leq\vartheta\leq1\right\}$ of distributions is decreasing in $\vartheta$, i.e. for a given $u\in{\cal X}$,
$$P_{\vartheta_1}\{\tr\leq u\}\geq P_{\vartheta_2}\{\tr\leq u\} \quad\hbox{for}\quad \vartheta_1<\vartheta_2.$$
It follows from that fact that the family of binomial distributions is decreasing in probability of a success and $P_\vartheta\{\tr= u\}$ is a convex combination of binomial distributions.

Let $\tr=u$ be observed. The (symmetric) confidence interval for $\vartheta$ at confidence level $\gamma$ based on the exact distribution of $\tr$ is $(\vartheta_L(u),\vartheta_U(u))$, where
$$\eqalign{
\vartheta_L(u)&=\cases{-1&for $u=-1$,\cr \max\left\{\vartheta: P_{\vartheta}\{\tr< u\}={1+\gamma\over2}\right\}&for $u>-1$,\cr}\cr
\vartheta_U(u)&=\cases{1&for $u=1$,\cr \min\left\{\vartheta: P_{\vartheta}\{\tr\leq u\}={1-\gamma\over2}\right\}&for $u<1$.\cr}\cr}\eqno{(M)}$$
%Here
%$$P_\theta\{\tr< u\}={1\over L(\theta)}\int_{a(\theta)}^{b(\theta)}\sum_{i_2=0}^{n_2}Q_{(\theta_1,n_1)}\left\{\xi_1\leq {n_1}\left(u+{i_2\over n_2}\right)-1\right\}Q_{(\theta_1-\theta,n_2)}\left\{\xi_2=i_2\right\}d\theta_1.$$
Unfortunately, closed formulae for such confidence intervals are not available. Nevertheless, for given $n_1$, $n_2$ and observed $u$ the confidence interval may be easily obtained with the standard mathematical software (for example R-project, Mathematica, MathLab etc.). Table 1 presents some $95\%$ confidence intervals for $n_1=n_2=10$ and $n_1=50$, $n_2=10$.

\midinsert
$$\vbox{\tabskip1em minus0.9em\offinterlineskip\halign{
\strut\hfil$#$\hfil&\hfil$#$\hfil&#\vrule&\hfil$#$\hfil&\hfil$#$\hfil&#\vrule\hskip2pt\vrule&
\hfil$#$\hfil&\hfil$#$\hfil&#\vrule&\hfil$#$\hfil&\hfil$#$\hfil\cr
\multispan{5}\strut{\bf Table 1.} Confidence intervals, $\gamma=0.95$\hfill\cr\noalign{\vskip5pt}
\multispan{5}\strut\hfill$n_1=n_2=10$\hfill\cr
\tr&interval&&\tr&interval\cr\noalign{\hrule}
-1.0	&	(-1.0000,-0.6733)	&&	0.1	&	(-0.3319,0.5171)	\cr
-0.9	&	(-0.9975,-0.5214)	&&	0.2	&	(-0.2326,0.6019)	\cr
-0.8	&	(-0.9751,-0.3940)	&&	0.3	&	(-0.1291,0.6813)	\cr
-0.7	&	(-0.9350,-0.2798)	&&	0.4	&	(-0.0212,0.7551)	\cr
-0.6	&	(-0.8832,-0.1745)	&&	0.5	&	(\m0.0760,0.8227)	\cr
-0.5	&	(-0.8227,-0.0760)	&&	0.6	&	(\m0.1745,0.8832)	\cr
-0.4	&	(-0.7551,\m0.0212)	&&	0.7	&	(\m0.2798,0.9350)	\cr
-0.3	&	(-0.6813,\m0.1291)	&&	0.8	&	(\m0.3940,0.9751)	\cr
-0.2	&	(-0.6019,\m0.2326)	&&	0.9	&	(\m0.5214,0.9975)	\cr
-0.1	&	(-0.5171,\m0.3319)	&&	1.0	&	(\m0.6733,1.0000)	\cr
\m0.0	&	(-0.4270,\m0.4270)	&&		&		                \cr\noalign{\vskip5pt}
\multispan{5}\hfill$n_1=50, n_2=10$\hfill\cr
\tr&interval&&\tr&interval\cr\noalign{\hrule}
	-1.0	&	(-1.0000,-0.8346)	&&	0.1	&	(-0.2103,0.4135)	\cr
	-0.9	&	(-0.9949,-0.6642)	&&	0.2	&	(-0.1046,0.5073)	\cr
	-0.8	&	(-0.9563,-0.5302)	&&	0.3	&	(\m0.0023,0.5962)	\cr
	-0.7	&	(-0.8986,-0.4105)	&&	0.4	&	(\m0.0971,0.6801)	\cr
-0.6	&	(-0.8322,-0.2998)	&&	0.5	&	(\m0.1957,0.7590)	\cr
-0.5	&	(-0.7590,-0.1957)	&&	0.6	&	(\m0.2998,0.8322)	\cr
	-0.4	&	(-0.6801,-0.0971)	&&	0.7	&	(\m0.4105,0.8986)	\cr
	-0.3	&	(-0.5962,-0.0023)	&&	0.8	&	(\m0.5302,0.9563)	\cr
	-0.2	&	(-0.5073,\m0.1046)	&&	0.9	&	(\m0.6642,0.9949)	\cr
	-0.1	&	(-0.4135,\m0.2103)	&&	1.0	&	(\m0.8346,1.0000)	\cr
\m0.0	&	(-0.3145,\m0.3145)	&&		&		\cr
}}$$
\endinsert

\setbox101=\vbox{\beginpicture
\setcoordinatesystem units <30truemm,58truecm>
\setplotarea x from -1 to 1, y from 0.94 to 1.00
\axis bottom
 ticks numbered from -1 to 1 by 0.50 /
\axis left
 ticks numbered from 0.94 to 1.00 by 0.01 /
\plot -1	1
-0.99	0.995595
-0.98	0.983368
-0.97	0.996959
-0.96	0.993215
-0.95	0.987526
-0.94	0.97971
-0.93	0.995716
-0.92	0.993121
-0.91	0.989628
-0.9	0.985123
-0.89	0.979504
-0.88	0.995004
-0.87	0.992974
-0.86	0.990411
-0.85	0.987245
-0.84	0.983409
-0.83	0.978843
-0.82	0.994253
-0.81	0.992496
-0.8	0.990366
-0.79	0.987818
-0.78	0.984806
-0.77	0.981287
-0.76	0.977217
-0.75	0.993269
-0.74	0.991624
-0.73	0.989685
-0.72	0.987421
-0.71	0.9848
-0.7	0.981788
-0.69	0.978353
-0.68	0.993342
-0.67	0.967968
-0.66	0.969228
-0.65	0.969907
-0.64	0.970015
-0.63	0.969559
-0.62	0.968542
-0.61	0.966963
-0.6	0.983433
-0.59	0.983331
-0.58	0.982892
-0.57	0.982115
-0.56	0.980993
-0.55	0.979521
-0.54	0.977689
-0.53	0.975486
-0.52	0.95139
-0.51	0.970279
-0.5	0.971486
-0.49	0.972259
-0.48	0.972608
-0.47	0.972538
-0.46	0.972053
-0.45	0.971154
-0.44	0.96984
-0.43	0.968108
-0.42	0.985067
-0.41	0.984721
-0.4	0.984124
-0.39	0.963942
-0.38	0.964923
-0.37	0.965448
-0.36	0.965525
-0.35	0.965159
-0.34	0.964355
-0.33	0.98092
-0.32	0.981211
-0.31	0.981224
-0.3	0.980962
-0.29	0.980428
-0.28	0.979621
-0.27	0.96135
-0.26	0.961858
-0.25	0.961914
-0.24	0.961524
-0.23	0.978502
-0.22	0.979088
-0.21	0.979383
-0.2	0.979393
-0.19	0.979124
-0.18	0.978579
-0.17	0.960224
-0.16	0.961025
-0.15	0.961376
-0.14	0.961287
-0.13	0.960765
-0.12	0.978738
-0.11	0.979221
-0.1	0.979424
-0.09	0.979352
-0.08	0.979011
-0.07	0.961708
-0.06	0.962655
-0.05	0.96317
-0.04	0.963264
-0.03	0.962947
-0.02	0.979886
-0.01	0.980472
0	0.980793
0.01	0.980472
0.02	0.979886
0.03	0.962947
0.04	0.963264
0.05	0.96317
0.06	0.962655
0.07	0.961708
0.08	0.979011
0.09	0.979352
0.1	0.979424
0.11	0.979221
0.12	0.978738
0.13	0.960765
0.14	0.961287
0.15	0.961376
0.16	0.961025
0.17	0.960224
0.18	0.978579
0.19	0.979124
0.2	0.979393
0.21	0.979383
0.22	0.979088
0.23	0.978502
0.24	0.961524
0.25	0.961914
0.26	0.961858
0.27	0.96135
0.28	0.979621
0.29	0.980428
0.3	0.980962
0.31	0.981224
0.32	0.981211
0.33	0.98092
0.34	0.964355
0.35	0.965159
0.36	0.965525
0.37	0.965448
0.38	0.964923
0.39	0.963942
0.4	0.984124
0.41	0.984721
0.42	0.985067
0.43	0.968108
0.44	0.96984
0.45	0.971154
0.46	0.972053
0.47	0.972538
0.48	0.972608
0.49	0.972259
0.5	0.971486
0.51	0.970279
0.52	0.95139
0.53	0.975486
0.54	0.977689
0.55	0.979521
0.56	0.980993
0.57	0.982115
0.58	0.982892
0.59	0.983331
0.6	0.983433
0.61	0.966963
0.62	0.968542
0.63	0.969559
0.64	0.970015
0.65	0.969907
0.66	0.969228
0.67	0.967968
0.68	0.993342
0.69	0.978353
0.7	0.981788
0.71	0.9848
0.72	0.987421
0.73	0.989685
0.74	0.991624
0.75	0.993269
0.76	0.977217
0.77	0.981287
0.78	0.984806
0.79	0.987818
0.8	0.990366
0.81	0.992496
0.82	0.994253
0.83	0.978843
0.84	0.983409
0.85	0.987245
0.86	0.990411
0.87	0.992974
0.88	0.995004
0.89	0.979504
0.9	0.985123
0.91	0.989628
0.92	0.993121
0.93	0.995716
0.94	0.97971
0.95	0.987526
0.96	0.993215
0.97	0.996959
0.98	0.983368
0.99	0.995595
1	1 /

\setdashpattern<1.5truemm,1.5truemm>
\plot -1 0.95 1 0.95 /
\put{$n_1=n_2=10$}  at 0.00 0.945
\put {{\bf Figure 1a.} Coverage probability of $(M)$.} at 0.00 0.925
\endpicture
}

\setbox102=\vbox{\beginpicture
\setcoordinatesystem units <30truemm,58truecm>
\setplotarea x from -1 to 1, y from 0.94 to 1.00
\axis bottom
 ticks numbered from -1 to 1 by 0.50 /
\axis left
 ticks numbered from 0.94 to 1.00 by 0.01 /
\plot -1	1
-0.99	0.991487
-0.98	0.989213
-0.97	0.984345
-0.96	0.976968
-0.95	0.984617
-0.94	0.975616
-0.93	0.979295
-0.92	0.980084
-0.91	0.97882
-0.9	0.97598
-0.89	0.980504
-0.88	0.983195
-0.87	0.977408
-0.86	0.979288
-0.85	0.979825
-0.84	0.979258
-0.83	0.95956
-0.82	0.958488
-0.81	0.964452
-0.8	0.962035
-0.79	0.966563
-0.78	0.957087
-0.77	0.961067
-0.76	0.959461
-0.75	0.955469
-0.74	0.961734
-0.73	0.960702
-0.72	0.957631
-0.71	0.956749
-0.7	0.954054
-0.69	0.958866
-0.68	0.964845
-0.67	0.963796
-0.66	0.957711
-0.65	0.957348
-0.64	0.962769
-0.63	0.952301
-0.62	0.957932
-0.61	0.962504
-0.6	0.953789
-0.59	0.960185
-0.58	0.952011
-0.57	0.958418
-0.56	0.950648
-0.55	0.956968
-0.54	0.962282
-0.53	0.952446
-0.52	0.957962
-0.51	0.957431
-0.5	0.955662
-0.49	0.955389
-0.48	0.961392
-0.47	0.953894
-0.46	0.959923
-0.45	0.952751
-0.44	0.958729
-0.43	0.957772
-0.42	0.957656
-0.41	0.955693
-0.4	0.95585
-0.39	0.961918
-0.38	0.95464
-0.37	0.960787
-0.36	0.95386
-0.35	0.960018
-0.34	0.95337
-0.33	0.959479
-0.32	0.953047
-0.31	0.959596
-0.3	0.959338
-0.29	0.958719
-0.28	0.958657
-0.27	0.958309
-0.26	0.958388
-0.25	0.958253
-0.24	0.958426
-0.23	0.958459
-0.22	0.95869
-0.21	0.958853
-0.2	0.959116
-0.19	0.952287
-0.18	0.959007
-0.17	0.952477
-0.16	0.959196
-0.15	0.952916
-0.14	0.959595
-0.13	0.953527
-0.12	0.960135
-0.11	0.95425
-0.1	0.960766
-0.09	0.954692
-0.08	0.961193
-0.07	0.955329
-0.06	0.961772
-0.05	0.956082
-0.04	0.962435
-0.03	0.956897
-0.02	0.950735
-0.01	0.957745
0	0.951684
0.01	0.957745
0.02	0.950735
0.03	0.956897
0.04	0.962435
0.05	0.956082
0.06	0.961772
0.07	0.955329
0.08	0.961193
0.09	0.954692
0.1	0.960766
0.11	0.95425
0.12	0.960135
0.13	0.953527
0.14	0.959595
0.15	0.952916
0.16	0.959196
0.17	0.952477
0.18	0.959007
0.19	0.952287
0.2	0.959116
0.21	0.958853
0.22	0.95869
0.23	0.958459
0.24	0.958426
0.25	0.958253
0.26	0.958388
0.27	0.958309
0.28	0.958657
0.29	0.958719
0.3	0.959338
0.31	0.959596
0.32	0.953047
0.33	0.959479
0.34	0.95337
0.35	0.960018
0.36	0.95386
0.37	0.960787
0.38	0.95464
0.39	0.961918
0.4	0.95585
0.41	0.955693
0.42	0.957656
0.43	0.957772
0.44	0.958729
0.45	0.952751
0.46	0.959923
0.47	0.953894
0.48	0.961392
0.49	0.955389
0.5	0.955662
0.51	0.957431
0.52	0.957962
0.53	0.952446
0.54	0.962282
0.55	0.956968
0.56	0.950648
0.57	0.958418
0.58	0.952011
0.59	0.960185
0.6	0.953789
0.61	0.962504
0.62	0.957932
0.63	0.952301
0.64	0.962769
0.65	0.957348
0.66	0.957711
0.67	0.963796
0.68	0.964845
0.69	0.958866
0.7	0.954054
0.71	0.956749
0.72	0.957631
0.73	0.960702
0.74	0.961734
0.75	0.955469
0.76	0.959461
0.77	0.961067
0.78	0.957087
0.79	0.966563
0.8	0.962035
0.81	0.964452
0.82	0.958488
0.83	0.95956
0.84	0.979258
0.85	0.979825
0.86	0.979288
0.87	0.977408
0.88	0.983195
0.89	0.980504
0.9	0.97598
0.91	0.97882
0.92	0.980084
0.93	0.979295
0.94	0.975616
0.95	0.984617
0.96	0.976968
0.97	0.984345
0.98	0.989213
0.99	0.991487
1	1 /
\setdashpattern<1.5truemm,1.5truemm>
\plot -1 0.95 1 0.95 /
\put{$n_1=50, n_2=10$}  at 0.00 0.945
\put {{\bf Figure 1b.} Coverage probability of $(M)$.} at 0.00 0.925
\endpicture
}

\midinsert
\centerline{\copy101\hss\copy102}
\endinsert
For a given $\vartheta\in(-1,1)$ the coverage probability, by construction, equals
$$\sum_{u=F^{-1}_\vartheta((1-\gamma)/2)}^{F^{-1}_\vartheta((1+\gamma)/2)}P_\vartheta\{\tr=u\},$$
where $F^{-1}_\vartheta(\cdot)$ is the quantile function of the distribution of $\tr$. Since the distribution of $\tr$ is discrete, the coverage probability is at least $\gamma$. Figure 1 shows the coverage probability of the confidence interval $(M)$ for $\gamma=0.95$ (the coverage probability is calculated not simulated).

The length of the confidence interval depends on the sample sizes $n_1$ and $n_2$. Suppose we may conduct $n$ trials including $n_1$ trials with success probability $\theta_1$ and $n_2=n-n_1$ trials with probability $\theta_2$. To find the optimal $n_1$, i.e. one minimizing the length, it is enough to minimize the distance between quantiles of orders ${1+\gamma\over2}$ and ${1-\gamma\over2}$ of the distribution of $\tr$. It is easy to note that the distribution of $\tr$ is unimodal, so it is enough to minimize the variance of $\tr$. This variance equals

$$D_\vartheta^2(\tr)={1\over L(\vartheta)}\int_{a(\vartheta)}^{b(\vartheta)}\left(D_{(\theta_1,n_1)}^2\left({\xi_1\over n_1}\right)+D_{(\theta_1-\vartheta,n_2)}^2\left({\xi_2\over n_2}\right)\right)d\theta_1=
{1-3\vartheta^2+2|\vartheta|^3\over6nf(1-f)},$$

where $f=n_1/n$. The variance $D_\vartheta^2(\tr)$ is (uniformly in $\vartheta$) minimal for $f=1/2$, i.e. half of the trials should be done with probability $\theta_1$. Hence, to obtain the maximal precision of estimation, i.e. the shortest (symmetric) confidence interval, the number of trials should be equally divided between the two groups. Of course this is possible in the case of a planned experiment. Unfortunately, in many real experiments (especially medical ones) it is not possible to have planned experiments.
\bigskip
\goodbreak
{\bf 3. A medical example}
\bigskip
The aim of the investigation was to compare the frequencies of occurrence of the specific immunoglobulin E G6 ({\it Phleum pratense} L.) in two sites: urban (represented by the Polish town Lublin) and rural (represented by the Polish district Zamoœæ). The investigation is part of the ECAP (ecap.pl/eng$\_$www/index$\_$home.html) project conducted by Prof. Boles³aw Samoliñski (Warsaw Medical University). The data are presented by his courtesy.

Let $\theta_t$ and $\theta_c$ denote the percentages of people with high concentration of sIgE G6 (at least $0.35$ IU/ml) in the town and in the country, respectively. We are interested in estimating the difference $\theta_t-\theta_c$ at confidence level $0.95$. A sample of size $n_t=743$ was drawn from the town, and a sample of size $n_c=329$ from the country.
%In that sample $k_t=90$ people have sIgE G6 concentrated at least $0.35$ IU/ml. From the country a sample of size $n_c=329$ was drawn and in that sample $k_c=25$ people have sIgE  G6 concentrated at least $0.35$ IU/ml.
The difference between the sample proportions equals $0.0603$. The confidence interval for the difference of proportions $\theta_t-\theta_c$ at confidence level $0.95$ is $(0.0052, 0.1154)$ (calculated from formula $(M)$ with $u=0.0603$). Since the lower end of the confidence interval is positive, we may conclude that the fraction of people with allergy to {\it Phleum pratense} L. is higher in the town than in the country.

In the above samples the level of the specific immunoglobulin E D1 ({\it Dermatophagoides pteronyssinus}) was also marked.
%In the urban region $m_t=42$ people were observed with sIgE D1 concentrated at least $0.35$ IU/ml, while in the country sample $m_c=9$ people have sIgE D1 concentrated at least $0.35$ IU/ml.
The question is the same as in the previous investigation: what is the difference between percentages of people with allergy to {\it Dermatophagoides pteronyssinus}  in urban and in rural areas. The difference between the observed proportions is $0.0292$ and confidence interval, at confidence level $0.95$, is $(-0.0276,0.0853)$. Since the confidence interval covers $0$, it may be supposed that the percentages of people with allergy to that allergen are the same.

%\vfill\eject
\bigskip
{\bf 4. Discussion and conclusions}
\bigskip
Estimating the difference of two binomial proportions is one of the crucial problems in medicine, biometrics etc. In this paper a new confidence interval for that difference is proposed. The confidence interval is based on the exact distribution of the sample difference, hence it works for large as well as for small samples. The coverage probability of that confidence interval is at least the nominal confidence level, in contrast to asymptotic confidence intervals known in the literature. It must be noted that the only information needed to construct the new confidence interval is sample sizes and sample difference between proportions, while for the confidence intervals appearing in the literature the knowledge of sample sizes as well as sample proportions in each sample is needed. Unfortunately it may lead to misunderstandings. Namely, suppose that seven experiments were conducted. In each experiment two samples of sizes fifty and ten respectively, were drawn ($n_1=50$, $n_2=10$). The resulting numbers of successes are shown in Table 2 (the first two columns).
\midinsert
$$\vbox{\tabskip1em minus0.9em\offinterlineskip\halign to\hsize{
\strut\hfil$#$\hfil&&#\vrule&\hfil$#$\hfil\cr
\multispan{11}\strut{\bf Table 2.} Confidence intervals in seven experiments\hfill\cr\noalign{\vskip5pt}
\xi_1&&\xi_2&&\tr&&\hbox{Wang c.i}&&\hbox{$K_1$ c.i}&&\hbox{$K_2$ c.i.}\cr\noalign{\hrule}
16&&0&&0.32&&(\m0.04738;0.47101)&&(\m0.01975;0.62025)&&(\m0.19070;0.44930)\cr
21&&1&&0.32&&(-0.00273;0.50696)&&(-0.00719;0.64719)&&(\m0.08915;0.55085)\cr
26&&2&&0.32&&(-0.03047;0.55617)&&(-0.01873;0.65873)&&(\m0.03602;0.60398)\cr
31&&3&&0.32&&(-0.02693;0.58380)&&(-0.01645;0.65645)&&(\m0.00571;0.63429)\cr
36&&4&&0.32&&(-0.02108;0.61329)&&(-0.00007;0.64007)&&(-0.00816;0.64816)\cr
41&&5&&0.32&&(\m0.00656;0.62735)&&(\m0.03283;0.60717)&&(-0.00769;0.64769)\cr
46&&6&&0.32&&(\m0.03955;0.63766)&&(\m0.08920;0.55080)&&(\m0.00718;0.63282)\cr
}}$$
\endinsert
It is seen that the sample difference between proportions (the third column) is the same in all experiments, but the confidence intervals are quite different (Table 2 gives results for three confidence intervals, but for other confidence intervals the results are similar). Moreover, for example application of $(K_1)$ or Wang confidence intervals in the sixth experiment suggests that $\tr=0.32$ is a statistically significant difference while in the fourth one it is not. The confidence interval $(M)$ we propose does not have this drawback: for observed $\tr$ we obtain one confidence interval whatever $\xi_1$ and $\xi_2$ are (here it is $(0.02110;0.61120)$).

Closed formulae for the new confidence interval are not available. But it is easy to calculate the confidence interval for given $n_1$, $n_2$ and an observed sample difference $\tr$ (see Appendix~1 for an exemplary R code). Because the proposed confidence interval may be applied for small as well as for large sample sizes, it may be recommended for practical use.

The coverage probability of the proposed confidence interval is at least the nominal confidence level. The equality of the coverage probability and the confidence level may be obtained by an appropriate randomization. The idea of randomized confidence intervals is presented for example in R. Zieliñski and W. Zieliñski (2005), W. Zieliñski (2014, 2017). The same idea may be applied to the proposed confidence interval; work on this is in progress.

%\vfill\eject
\bigskip
{\bf References}

\def\pn#1{\noindent\hangindent=1em \hangafter=1 #1\par}

\pn{Anbar, D. (1983) On Estimating the Difference between Two Probabilities with Special Reference to Clinical Trials, Biometrics, 39, 257-262.}

\pn{Beal, S. L. (1987) Asymptotic Confidence Intervals for the Difference between Two Binomial Parameters for Use with Small Samples, Biometrics, 43, 941-950.}

\pn{Brown, L. D., Cai, T. T., DasGupta A. (2001) Interval Estimation for a Binomial Proportion, Statistical Science, 16, 101-133.}

\pn{Fleiss, J. L.  (1981) Statistical Methods for Rates and Proportions, 2nd ed., Wiley, New York.}

\pn{Mee, R. W. and Anbar, D. (1984) Confidence Bounds for the Difference between Two Probabilities. Biometrics 40, 1175-176.}

\pn{Miettinen, O. S. and Nurminen, M. (1985) Comparative Analysis of Two Rates, Statistics in Medicine, 4, 213-226.}

\pn{Newcombe, R. (1998) Interval Estimation for The Difference Between Independent Proportions: Comparison of Eleven Methods, Statistics in Medicine, 17, 873-890.}

\pn{Neyman, J. (1934) On the Two Different Aspects of the Representative Method: The Method of Stratified Sampling and the Method of Purposive Selection, Journal of the Royal Statistical Society, 97, 558-625.}

\pn{Shan, G. and Wang, W. (2013) ExactCIdiff: An R Package for Computing Exact Confidence Intervals for the Difference of Two Proportions, The R Journal Vol. 5/2, 62-70.}

\pn{Wang, W. (2010) On Construction of the Smallest One-Sided Confidence Interval for the Difference of Two Proportions, The Annals of Statistics, 38, 1227-1243, doi: 10.1214/09\=AOS744}

\pn{Wilson, E. B. (1927) Probable inference, the law of succession, and statistical inference, Journal of the American Statistical Association, 22, 209-212.}

\pn{Zhou, X-H., Tsao, M. and Qin, G. (2004) New intervals for the difference between two independent binomial proportions. J. Statist. Plann. Inference 123, 97-115,\hfill\break doi: 10.1016/S0378-3758(03)00146-0.}

\pn{Zieliñski, R. and Zieliñski, W. (2005) Best Exact Nonparametric Confidence Intervals for Quantiles, Statistics 39, 67-71, doi: 10.1080/02331880412331329854.}

\pn{Zieliñski, W. (2014) The shortest randomized confidence interval for probability of success in a negative binomial model, Applicationes Mathematicae 41, 43-49,\hfill\break doi: 10.4064/am41-1-4}

\pn{Zieliñski, W. (2017) The shortest Clopper-Pearson randomized confidence interval for binomial probability, REVSTAT-Statistical Journal	15, 141-153.}

\pn{Zieliñski, W. (2018) Confidence Interval for the Weighted Sum of Two Binomial Proportions, Applicationes Mathematicae 45, 53-60.}

\vfill\eject

\bigskip
{\bf Appendix 1}
\bigskip
An exemplary R code for calculating the confidence interval is enclosed. I~am grateful to Prof. Stanis³aw Jaworski for his help.
\bigskip

\begingroup
\font\ttt=pltt8
\ttt
\obeylines
\baselineskip=0.4\baselineskip

CI=function(uemp,n,gamma)$\{$
  u=abs(uemp)
  g=function(u,vartheta,lq=0)$\{$
    f=function(theta,k)$\{$pbinom(n[1]*(u+k/n[2])-lq,n[1],theta)*dbinom(k,n[2],theta-vartheta)$\}$
    a=max(0,vartheta)
    b=min(1,1+vartheta)
    wynik=c()
    for (k in 0:n[1])$\{$wynik[k+1]=integrate(f,a,b,k=k)\$value $\}$
    t=sum(wynik)/(b-a)
    (t-(1+gamma*(-1+2*lq))/2)\^{}2$\}$
  P=ifelse(u==1,1,optimize(g,c(u,1),u=u)\$minimum) \# upper
  L=optimize(g,c(-1,u),u=u,lq=1)\$minimum \# lower
  info=paste("at 1-alpha=",gamma,", where u=",uemp, ", n1=",n[1],", n2=",n[2],sep="")
  if (uemp>0)
  $\{$paste("Confidence interval (",round(L,4),",",round(P,4),") ",info,sep="")$\}$
  else
  $\{$paste("Confidence interval (",round(-P,4),",",round(-L,4),") ",info,sep="")$\}$
$\}$

\#Example of usage
  n=c(10,10) \# input n1 and n2
  CI(-0.3,n,gamma=0.99) \# input the observed difference and the confidence level
\endgroup

\vfill\eject

\bigskip
{\bf Appendix 2}
\bigskip

Confidence intervals for $\vartheta=\theta_1-\theta_2$ appearing in the literature are constructed for ``large'' sample sizes $n_1$ and $n_2$. It is assumed that $\xi_1$ and $\xi_2$ (and so $\xi_1-\xi_2$) are normally distributed. In what follows, $\gamma$ denotes the assumed confidence level and $z=z_{(1+\gamma)/2}$ denotes the quantile of order $(1+\gamma)/2$ of the standard normal distribution.

{\bf 1.} The approximate confidence interval based on the test statistic of the hypothesis $H:\theta_1=\theta_2$ has the form
$$\tr\pm z\sqrt{{\xi_1+\xi_2\over n_1+n_2}\left(1-{\xi_1+\xi_2\over n_1+n_2}\right)\left({1\over n_1}+{1\over n_2}\right)}.\eqno{(K_1)}$$
This is one of the most common confidence intervals. It may be found in various statistical textbooks (for example  https://onlinecourses.science.psu.edu/stat414/node/268).

{\bf 2.}  By the de Moivre-Laplace theorem, asymptotically $\tr\sim N\left(\theta,{\theta_1(1-\theta_1)\over n_1}+{\theta_2(1-\theta_2)\over n_2}\right)$. A~simple application of the asymptotic distribution gives $$\tr\pm z\sqrt{{\hat\theta_1(1-\hat\theta_1)\over n_1}+{\hat\theta_2(1-\hat\theta_2)\over n_2}}\eqno{(K_2)}$$
(for example stattrek.com/estimation/difference-in-proportions.aspx?Tutorial=AP). Mee and Anbar (1984) expressed the above interval in terms of $\tr$:
$$\tr\pm z\sqrt{{(\tilde\psi+\tr/2)(1-\tilde\psi-\tr/2)\over n_1}+{(\tilde\psi-\tr/2)(1-\tilde\psi+\tr/2)\over n_2}},$$
where $\tilde\psi=(\hat\theta_1+\hat\theta_2)/2$.

Miettinen and Nurminen (1985) slightly modified the above confidence interval:
$$\tr\pm z\sqrt{{n_1+n_2\over n_1+n_2-1}\left\{{(\tilde\psi+\tr/2)(1-\tilde\psi-\tr/2)\over n_1}+{(\tilde\psi-\tr/2)(1-\tilde\psi+\tr/2)\over n_2}\right\}}.\eqno{(K_2^\prime)}$$

{\bf 3.} The binomial distribution is a discrete one and is approximated by a continuous distribution. Hence the so called continuity correction is introduced (Fleiss 1981, p. 29): $$\tr\pm z\sqrt{{\xi_1(n_1-\xi_1)\over n_1^3}+{\xi_2(n_2-\xi_2)\over n_2^3}+{1\over2}\left({1\over n_1}+{1\over n_2}\right)}.\eqno{(K_3)}$$
This confidence interval is very conservative: its coverage probability is significantly higher than the assumed confidence level (see Figure 4).

{\bf 4.} Using the Haldane method, Beal (1987) obtained the confidence interval
$$\vartheta^*\pm w,\eqno{(K_4)}$$ where
$$\eqalign{
\vartheta^*&={\tr+z^2\nu(1-2\tilde\psi)\over1+z^2u},\cr
w&={z\over1+z^2u}\sqrt{u\{4\tilde\psi(1-\tilde\psi)-\tr^2\}+2\nu(1-2\tilde\psi)\tr+4z^2u^2(1-\tilde\psi)\tilde\psi+z^2\nu^2(1-2\tilde\psi)^2},\cr
\tilde\psi&={1\over2}\left(\hat\theta_1+\hat\theta_2\right)\quad
u={1\over4}\left({1\over n_1}+{1\over n_2}\right)\quad
\nu={1\over4}\left({1\over n_1}-{1\over n_2}\right).\cr
}$$
Using the Jeffreys-Perks method he obtained a similar confidence interval with
$$\tilde\psi={1\over2}\left({\xi_1+0.5\over n_1+1}+{\xi_2+0.5\over n_2+1}\right).\eqno{(K_4^\prime)}$$

{\bf 5.} The method based on the Wilson (1927) score method for the single proportion gives the confidence interval
$$L=\tr-\delta_{12},\quad U=\tr+\delta_{21},\eqno{(K_5)}$$
where
$$\delta_{ij}=\sqrt{(\hat\theta_i-l_i)^2+(u_j-\hat\theta_j)^2}=z\sqrt{l_i(1-l_i)/n_i+u_j(1-u_j)/n_j}$$
and $l_i$ and $u_i$ are the roots of $|\hat\theta_i-\theta_i|=z\sqrt{\theta_i(1-\theta_i)/n_i}$. Note that $l_i=0$ for $\xi_i=0$ and $u_i=1$ for $\xi_i=n_i$.

Using the continuity-correction score intervals, Fleiss (1981, pp. 13-14) obtained $l_i$ and $u_i$ as the solutions of
$$\left|\hat\theta_i-\theta_i\right|-{1\over2n_i}=z\sqrt{\theta_i(1-\theta_i)\over n_i}.\eqno{(K_5^\prime)}$$

{\bf 6.} Zhou et al. (2004) proposed two new confidence intervals based on the asymptotic Edgworth expansion of $\hat\theta_1-\hat\theta_2$. The first one is
$$\left(\tr-{\hat\sigma\over\sqrt{n}}\left(z-{{\hat Q}(z)\over\sqrt{n}}\right),
\tr+{\hat\sigma\over\sqrt{n}}\left(z+{{\hat Q}(z)\over\sqrt{n}}\right)\right),\eqno{(K_6)}$$
where ($n=n_1+n_2$)
$${\hat Q}(t)={{\hat a}+{\hat b}t^2\over\hat\sigma},\ {\hat\sigma}=\sqrt{n}\sqrt{{\xi_1(n_1-\xi_1)\over n_1^3}+{\xi_2(n_2-\xi_2)\over n_2^3}},\ {\hat a}={\hat\delta\over6{\hat\sigma}^2},
\ {\hat b}={n(n_1-2\xi_1)\over2n_1^2}-{\hat a},$$
$${\hat\delta}=\left({n\over n_1}\right)^2{\xi_1(n_1-\xi_1)(n_1-2\xi_1)\over n_1^3}-\left({n\over n_2}\right)^2{\xi_2(n_2-\xi_2)(n_2-2\xi_2)\over n_2^3}.$$
The second confidence interval has the form
$$\left(\tr-{\hat\sigma\over\sqrt{n}}g^{-1}(z),
\tr-{\hat\sigma\over\sqrt{n}}g^{-1}(-z)\right),\eqno{(K_7)}$$
where
$$g^{-1}(u)={\sqrt{n}\over{\hat b}{\hat\sigma}}\left(\left(1+3({\hat b}{\hat\sigma})\left({u\over\sqrt{n}}-{{\hat a}{\hat\sigma}\over n}\right)\right)^{1/3}-1\right).$$

The upper ends of the above mentioned confidence intervals may be greater than one (or their lower ends may be smaller than $-1$). It is customary to truncate such an interval at $1$ (or $-1$ respectively), but such an operation results in a very low coverage probability for values of $\vartheta$ near $1$ (or $-1$ respectively).

Wang (2010) (see also Shan and Wang, 2013) proposed a confidence interval which does not have the above disadvantage.

\bye